\documentclass[preprint2]{aastex}

\newcommand{\kms}{\ensuremath{\rm{km\,s}^{-1}}}
\newcommand{\msun}{\ensuremath{\rm{M}_\odot}}
\newcommand{\rsun}{\ensuremath{\rm{R}_{\odot}}}
\newcommand{\mjup}{\ensuremath{\rm{M}_{\rm Jup}}}
\newcommand{\rjup}{\ensuremath{\rm{R}_{\rm Jup}}}

\newcommand{\vsini}{$v \sin i$}

\newcommand{\halpha}{\mbox{$H_\alpha$}}
\newcommand{\hbeta}{\mbox{$H_\beta$}}
\newcommand{\teff}{\mbox{$T_{\rm eff}$}}
\newcommand{\logg}{\mbox{$\log g$}}
\newcommand{\mictrb}{\mbox{$\xi_{\rm t}$}}
\newcommand{\mactrb}{\mbox{$v_{\rm mac}$}}

\newcommand{\dchs}{\hbox{$\Delta\chi^2$}}

\slugcomment{Submitted to Astrophysical Journal}
\shorttitle{WASP-16b}
\shortauthors{Lister et al.}

\begin{document}

\title{WASP-16b: A new Jupiter-like planet transiting a southern solar analog}

\author{T.~A.~Lister\altaffilmark{1}, 
D.~R.~Anderson\altaffilmark{2}, 
M.~Gillon\altaffilmark{3,4}, 
L.~Hebb\altaffilmark{5}, 
B.~S.~Smalley\altaffilmark{2},
A.~H.~M.~J.~Triaud\altaffilmark{3},
A.~Collier~Cameron\altaffilmark{5},
D.~M.~Wilson\altaffilmark{2,6}, 
R.~G.~West\altaffilmark{7},
S.~J.~Bentley\altaffilmark{2}, 
D.~J.~Christian\altaffilmark{8,9}, 
R.~Enoch\altaffilmark{10,5}, 
C.~A.~Haswell\altaffilmark{10}, 
C.~Hellier\altaffilmark{2},
K.~Horne\altaffilmark{5},
J.~Irwin\altaffilmark{11}, 
Y.~C.~Joshi\altaffilmark{8}, 
S.~R.~Kane\altaffilmark{12}, 
M.~Mayor\altaffilmark{3}, 
P.~F.~L.~Maxted\altaffilmark{2},
A.~J.~Norton\altaffilmark{10}, 
N.~Parley\altaffilmark{10,5}, 
F.~Pepe\altaffilmark{3}, 
D.~Pollacco\altaffilmark{8}, 
D.~Queloz\altaffilmark{3}, 
R.~Ryans\altaffilmark{8}, 
D.~Segransan\altaffilmark{3}, 
I.~Skillen\altaffilmark{13}, 
R.~A.~Street\altaffilmark{1},
I.~Todd\altaffilmark{8}
S.~Udry\altaffilmark{3}, 
P.~J.~Wheatley\altaffilmark{14}
}
\email{tlister@lcogt.net}
\altaffiltext{1}{Las Cumbres Observatory, 6740 Cortona Drive Suite 102, Goleta,
CA 93117, USA}
\altaffiltext{2}{Astrophysics Group, Keele University, Staffordshire, ST5 5BG, UK}
\altaffiltext{3}{Observatoire de Gen\`eve, Universit' de Gen\`eve, 51 Ch.
des Maillettes, 1290 Sauverny, Switzerland}
\altaffiltext{4}{Institut d'Astrophysique et de G\'eophysique,  Universit\'e de Li\`ege,  All\'ee du 6 Ao\^ut, 17,  Bat.  B5C, Li\`ege 1, Belgium}
\altaffiltext{5}{SUPA, School of Physics and Astronomy, University of St Andrews,
North Haugh, St Andrews, Fife KY16 9SS, UK}
\altaffiltext{6}{Centre for Astrophysics \& Planetary Science, School of Physical
Sciences, University of Kent, Canterbury, Kent, CT2 7NH, UK}
\altaffiltext{7}{Department of Physics and Astronomy, University of Leicester,
Leicester, LE1 7RH, UK}
\altaffiltext{8}{Astrophysics Research Centre, School of Mathematics \& Physics,
Queen's University, University Road, Belfast, BT7 1NN, UK}
\altaffiltext{9}{California State University Northridge 18111 Nordhoff Street,
Northridge, CA 91330-8268, USA}
\altaffiltext{10}{Department of Physics and Astronomy, The Open University, Milton
   Keynes, MK7 6AA, UK}
\altaffiltext{11}{Harvard-Smithsonian Center for Astrophysics, 60 Garden Street,
Cambridge, MA, 02138 USA}
\altaffiltext{12}{NASA Exoplanet Science Institute, Caltech, MS 100-22, 770 South Wilson Avenue, Pasadena, CA 91125, USA}
\altaffiltext{13}{Isaac Newton Group of Telescopes, Apartado de Correos 321,
E-38700 Santa Cruz de la Palma, Tenerife, Spain}
\altaffiltext{14}{Department of Physics,   University of Warwick, Coventry CV4
7AL, UK}
\addtocounter{footnote}{14}
\begin{abstract}
We report the discovery from WASP-South of a new Jupiter-like extrasolar planet,
WASP-16b, which transits its solar analog host star every 3.12 days. Analysis of
the transit photometry and radial velocity spectroscopic data leads to a planet
with $R_{\rm p}=1.008\pm0.071$ \rjup\ and $M_{\rm p}=0.855\pm0.059$ \mjup,
orbiting a host star with  $R_*=0.946\pm0.054$ \rsun\ and
$M_*=1.022\pm0.101$ \msun. Comparison of the high resolution stellar spectrum
with synthetic spectra and stellar evolution models indicates the host star is a
near-solar metallicity ({[Fe/H]}$=0.01 \pm 0.10$) solar analog (\teff$=5700 \pm
150$ K, \logg$=4.5 \pm 0.2$) of intermediate age
($\tau=2.3^{+5.8}_{-2.2}\,\rm{Gyr}$).
\end{abstract}


\section{Introduction}

There are currently over 300 known
exoplanets\footnote{\url{http://exoplanet.eu}} with the majority of them
discovered through the radial velocity technique. A growing number of exoplanets
in recent years have been discovered through the transit method. Transiting
exoplanets are particularly valuable as they allow parameters such as the mass,
radius and density to be accurately determined and further studies such as
transmission spectroscopy, secondary eclipse measurements and transit timing
variations to be carried out.

There are several wide angle surveys that have been successful in finding
transiting exoplanets around bright stars, namely HAT \citep{bakos2002hat}, TrES
\citep{alonso2004}, XO \citep{mccullough2005} and WASP \citep{pollacco2006}. The
WASP Consortium conducts the only exoplanet search currently operating in both
hemispheres although HATnet is planning a southern extension and several groups
are planning searches from Antarctica (e.g.
\citealt{strassmeier2007icet,crouzet2008}).

We report the discovery from the WASP-South observatory of a $\sim0.86$ \mjup\ mass
companion orbiting a $V\sim11.3$ close solar analog WASP-16 (=TYC 6147-229-1, USNO-B1.0
0697-0298329). 

\section{Observations}

\subsection{Photometric observations}

WASP-South, located at SAAO, South Africa, is one of two SuperWASP instruments
and comprises eight cameras on a robotic mount. Each camera consists of
a Canon 200mm f/1.8 lens with an Andor $2048\times2048$ e2v CCD camera giving a
field of view of $7.8\arcdeg \times 7.8\arcdeg$ and a pixel scale of
13.7\arcsec. Exposure times were 30\,s and the same field is returned to and
reimaged every 8--10 minutes. Further details of the instrument, survey and 
data reduction pipelines are given in \citet{pollacco2006} and the candidate
selection procedure is described in \citet{cameron2007} and
\citet{pollacco2008wasp3} and references therein.

WASP-16 was observed for a partial season in 2006, a full season in 2007 and a further
partial season in 2008 with the distribution of data points as 3324 points (2006), 6013
(2007) and 4084 (2008). The 2007 light curve revealed the presence of a $\sim1.3\%$ dip
with a period of $\sim3.11$\,days. The transit coverage in the other two seasons was very
sparse, particularly in 2006, and there is only evidence for 2 partial transits in the
2008 data. WASP-16 was a fairly strong candidate for follow-up despite the small number
of transits, passing the filtering tests of \citet{cameron2006} with a signal to red
noise ratio, $S_{\mathrm{red}}=9.38$ (with $S_{\mathrm{red}}>5$ required for selection),
`transit to antitransit ratio' \dchs/\dchs$_- = 2.5$ (\dchs/\dchs$_- \geq 1.5$ required
for selection) and no measurable ellipsoidal variation.

\begin{figure}
\epsscale{.8}
\plotone{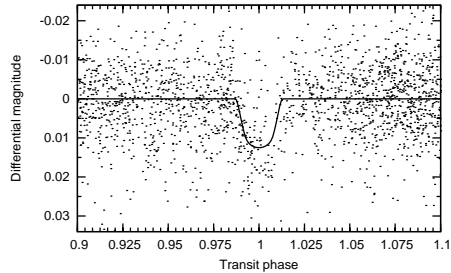}
\caption{Zoom of the transit region of the SuperWASP light curve of WASP-16b
with the best fitting MCMC model overplotted}
\protect\label{fig:lc_SWASP_zoom}
\end{figure}

\begin{figure}
\epsscale{.8}
\plotone{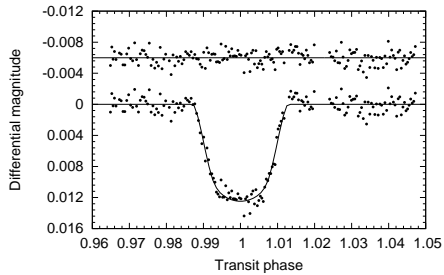}
\caption{$I_c$ band light curve from EULERCAM and residuals from the transit fit of WASP-16b}
\protect\label{fig:lc_EULER_I}
\end{figure}

The SuperWASP light curve showing a zoom of the transit region, along with the
model transit fit, is shown in Figure~\ref{fig:lc_SWASP_zoom}. In order to
better constrain the transit parameters, follow-up high precision photometric
observations with the Swiss 1.2m+EULERCAM on La Silla, were obtained in the
$I_c$ band on the night of 2008 May 04 and are shown in
Figure~\ref{fig:lc_EULER_I}.

\subsection{Spectroscopic observations}

\begin{figure}
\epsscale{.7}
\plotone{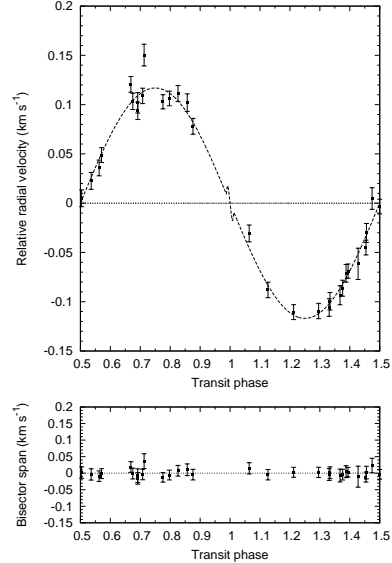}
\caption{Radial velocity curve (upper panel) of WASP-16 from the Swiss
1.2m+CORALIE along with the best-fitting model which includes the predicted
Rossiter-McLaughlin effect. The resulting bisector spans are shown in the lower
panel. The uncertainties on the bisector spans are double the radial velocity
uncertainties.}
\protect\label{fig:RV}
\end{figure}

In order to confirm the planetary nature of the transit signal, we obtained
follow-up spectroscopic observation with the Swiss 1.2m+CORALIE spectrograph.
The data were processed through the standard CORALIE reduction pipeline as
described by \citet{baranne1996} with an additional correction for the blaze
function. Fourteen radial velocity measurements were made between 2008 March 10
and 2008 August 04 and an additional sixteen between 2009 Feb 19 and 2009 June
03 (see Table~\ref{tab:wasp16rvs}) by cross-correlating with a G2 template mask.
The resulting radial velocity (RV) curve is shown in Figure~\ref{fig:RV}. The
low amplitude RV variation clearly supports the existence of a planetary mass
companion. In order to rule out a non-planetary explanation for the radial
velocity variation such as a blended eclipsing binary or starspots, we examined
the line-bisector spans. Contamination from an unresolved eclipsing binary will
cause asymmetries in the spectral line profiles and line bisector span
variations \citep{queloz2001,torres2005}. As can be seen from the lower panel of
Figure~\ref{fig:RV}, there is no sign of variation with phase of the bisector
spans and their amplitude is much smaller than the radial velocity variation.
This supports the conclusion that the radial velocity variations are due to a
planet orbiting the star and not some other cause.

\begin{deluxetable}{cccc}
\tablecaption{CORALIE radial velocities for WASP-16.\label{tab:wasp16rvs}}
\tablehead{
\colhead{Time of obs.} & \colhead{Rad. Vel.} & \colhead{$\sigma_{RV}$} & \colhead{Bisector span} \\
\colhead{(BJD-2450000)} & \colhead{(\kms)} & \colhead{(\kms)} & \colhead{(\kms)}}
\tablewidth{0pt}
\tablecolumns{4}
\startdata
4535.864842 & -1.99772 & 0.01591 & 0.00306 \\
4537.849158 & -1.96688 & 0.00853 & -0.04553 \\
4538.858364 & -2.00734 & 0.00899 & -0.03129 \\
4558.780835 & -1.83336 & 0.00723 & -0.02779 \\
4560.709473 & -2.00513 & 0.00725 & -0.02403 \\
4561.688137 & -1.82730 & 0.00785 & -0.03998 \\
4589.705102 & -1.84255 & 0.00875 & -0.04520 \\
4591.706755 & -2.03571 & 0.00892 & -0.03221 \\
4652.495906 & -1.82493 & 0.00808 & -0.03209 \\
4656.551645 & -2.02421 & 0.00787 & -0.02555 \\
4657.577293 & -1.96640 & 0.00957 & -0.01827 \\
4663.539741 & -2.02961 & 0.00969 & -0.02661 \\
4664.616769 & -1.78590 & 0.01108 & -0.04350 \\
4682.521501 & -1.98118 & 0.00754 & -0.02123 \\
4881.869213 & -2.02245 & 0.00813 & -0.02760 \\
4882.801025 & -1.83289 & 0.00823 & -0.03739 \\
4884.737094 & -2.04565 & 0.00778 & -0.01672 \\
4891.805707 & -1.90043 & 0.00798 & -0.01009 \\
4892.723980 & -1.83413 & 0.00891 & -0.02116 \\
4941.728231 & -1.88737 & 0.00748 & -0.04134 \\
4943.730102 & -2.04677 & 0.00753 & -0.01825 \\
4944.739293 & -1.91359 & 0.00860 & -0.02245 \\
4945.799895 & -1.85815 & 0.00807 & 0.01502 \\
4947.745317 & -1.93960 & 0.00741 & -0.03134 \\
4948.673112 & -1.82992 & 0.00743 & -0.06231 \\
4972.707323 & -1.93123 & 0.00854 & -0.03631 \\
4975.733486 & -1.93144 & 0.01100 & -0.01416 \\
4982.647535 & -1.83433 & 0.01036 & -0.02677 \\
4984.642389 & -2.04210 & 0.00892 & -0.04270 \\
4985.694776 & -1.81561 & 0.00802 & -0.02406 \\

\enddata
\end{deluxetable}

\section{WASP-16 System Parameters}

\subsection{Stellar Parameters}

The individual CORALIE spectra are of relatively low signal-to-noise, but when
co-added into 0.01\AA\ steps they give a S/N of around 70:1 which is suitable
for a photospheric analysis of the host star. In addition, a single HARPS
spectrum was used to complement the CORALIE analysis, but this spectrum had
relatively modest S/N of around 50:1. The standard CORALIE/HARPS pipeline
reduction products were used in the analysis.

The analysis was performed in a very similar fashion to that described by
\cite{west2009wasp15} using a spectral synthesis package and LTE model
atmospheres. The \halpha\ and \hbeta\ lines were used to determine the effective
temperature (\teff), while the Na {\sc i} D and Mg {\sc i} b lines were used as
surface gravity (\logg) diagnostics. In addition the Ca H \& K lines provided a
further check on the derived \teff\ and \logg. The elemental abundances of
several elements were determined from measurements of several clean and
unblended lines. The parameters and abundances obtained from the analysis are
listed in Table~\ref{wasp16-params}.

\begin{deluxetable}{cc}
\tablecaption{Stellar parameters of the WASP-16 host star.\label{wasp16-params}}
\tablehead{
\colhead{Parameter} & \colhead{Value}}
\tablewidth{0pt}
\tablecolumns{2}
\startdata
\multicolumn{2}{l}{R.A. = 14$^{\rm h}$18$^{\rm m}$43\fs 92, Dec = -20\degr
16\arcmin 31\farcs 8 (J2000.0)} \\
\teff      & 5700 $\pm$ 150 K \\
\logg      & 4.5 $\pm$ 0.2 \\
\mictrb    & 1.1 $\pm$ 0.2 \kms \\
\vsini     & 3.0 $\pm$ 1.0 \kms \\
Spectral Type & G3V\tablenotemark{a} \\
{[Fe/H]}   & 0.01 $\pm$ 0.10 \\
{[Na/H]}   & 0.15 $\pm$ 0.08 \\
{[Mg/H]}   & 0.14 $\pm$ 0.10 \\
{[Si/H]}   & 0.10 $\pm$ 0.07 \\
{[Ca/H]}   & 0.11 $\pm$ 0.12 \\
{[Sc/H]}   & 0.14 $\pm$ 0.07 \\
{[Ti/H]}   & 0.05 $\pm$ 0.14 \\
{[V/H]}    & 0.09 $\pm$ 0.15 \\
{[Cr/H]}   & 0.02 $\pm$ 0.11 \\ 
{[Co/H]}   & 0.17 $\pm$ 0.08 \\
{[Ni/H]}   & 0.07 $\pm$ 0.12 \\
$\log A(\rm{Li})$& $<$ 0.8 \\
\teff(IRFM)&5550 $\pm$ 130 K\\
$\theta$(IRFM)& 0.052 $\pm$ 0.003 mas \\
\enddata
\tablenotetext{a}{Estimated from J-H color} \\
\end{deluxetable}

In our spectra the Li {\sc i} 6708\AA\ line is not detected (EW $<$ 2m\AA),
allowing us to derive an upper-limit on the Lithium abundance of log n(Li/H) +
12 $<$ 0.8. The lack of lithium would imply an age in excess of 5\,Gyr
\citep{sestito2005}. The stellar rotation velocity (\vsini) was determined by
fitting the profiles of several Fe~{\sc i} lines using an average value of
\mactrb\ = 2.0 \kms\ for the macroturbulence (\mactrb).

In addition to the spectral analysis, we have also used available broad-band
photometry to estimate the total observed bolometric flux. The Infrared Flux
Method \citep{blackwell1977} was then used with 2MASS magnitudes to determine
\teff\ and stellar angular diameter ($\theta$). This gives \teff = 5550 $\pm$
130~K, which is in close agreement with that obtained from the spectroscopic
analysis (\teff = 5700 $\pm$ 150~K).

\begin{figure}
\epsscale{1.0}
\plottwo{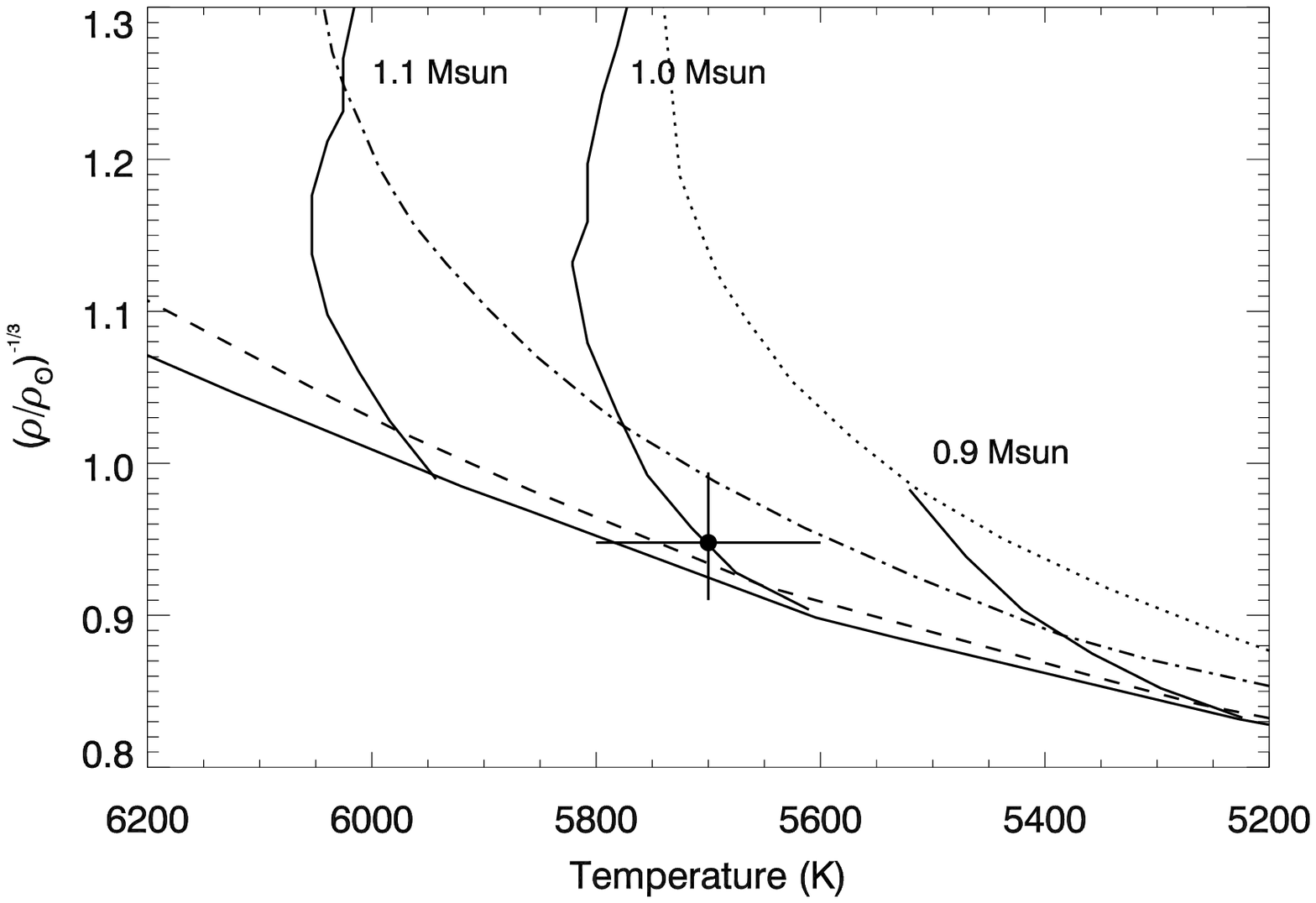}{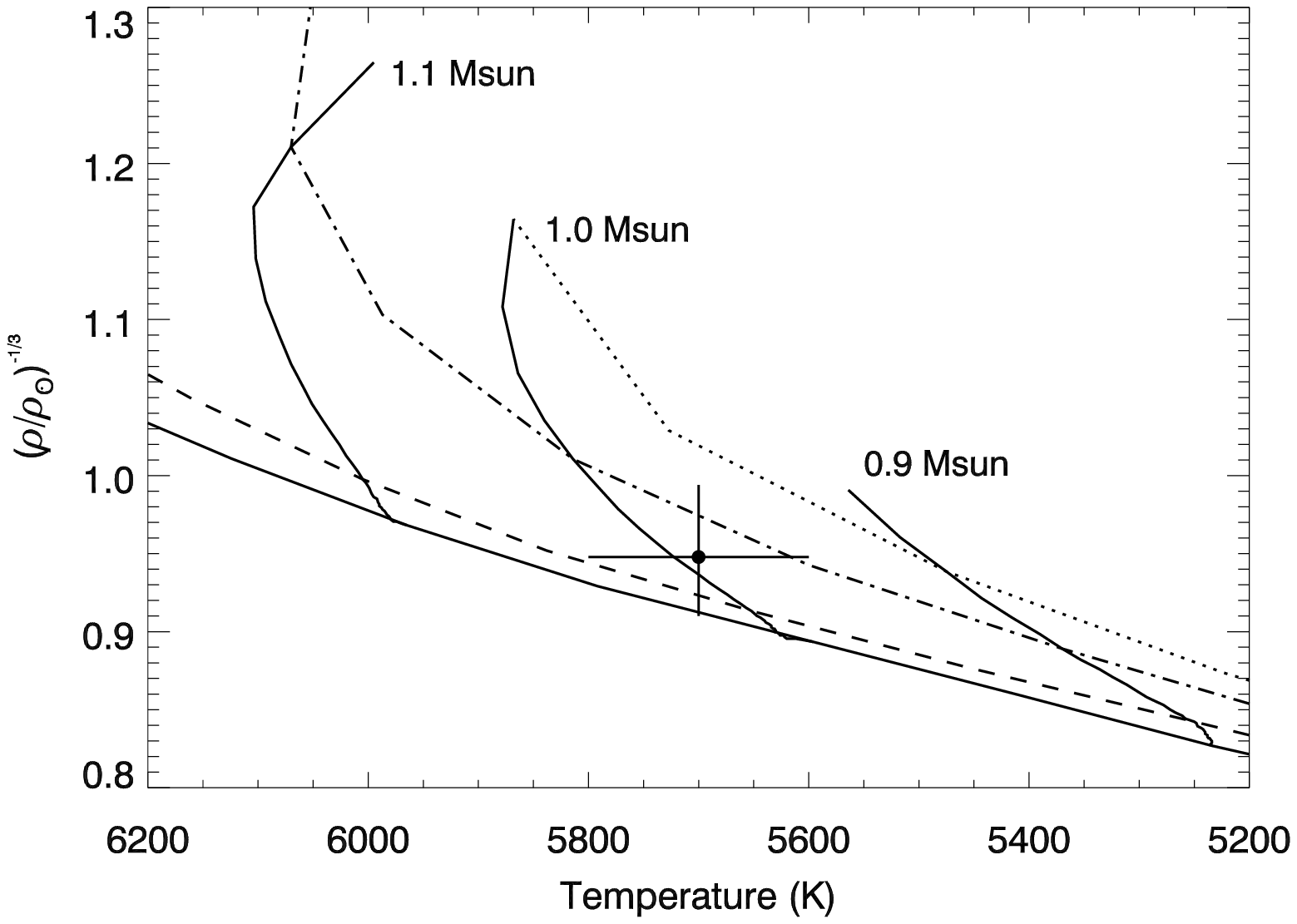}
\caption{The position of the WASP-16 host star in the isochrones of
\cite{girardi2000} (left) and \cite{baraffe1998} (right). In both plots the 0.9, 1.0
and 1.1 \msun mass tracks are shown along with 100 Myr (solid), 1 Gyr (dashed),
5 Gyr (dot-dashed) and 10 Gyr (dotted) isochrones.}
\protect\label{fig:isochrones}
\end{figure}

Comparison with the stellar evolution models of \citet{girardi2000} for solar
metallicity ($Z=0.02$) gives maximum-likelihood values $M_* =
1.00^{+0.045}_{-0.067} \msun$ 
as shown in Figure~\ref{fig:isochrones}. Alternative models from
\citet{baraffe1998} give essentially the same results as the stellar evolution
models have close agreement in this mass range. The uncertainties on the stellar
density lead to a large uncertainty on the age from the \citet{girardi2000}
isochrones producing an estimated age of $\tau=2.3^{+5.8}_{-2.2}\,\rm{Gyr}$.

\begin{deluxetable}{ccc}
\tablecaption{System parameters for WASP-16b.\label{tab:wasp16-sysparams}}
\tablehead{
\colhead{Parameter} & \colhead{Value} & \colhead{Error}}
\tablewidth{0pt}
\tablecolumns{3}
\startdata
$P$ (days)  & 3.1186009  & $^{+0.0000146}_{-0.0000131}$ \\
$T_0$ (HJD) & 2454584.42878 & $^{+0.00035}_{-0.00023}$ \\
$T_{dur}$ (days) & 0.0800 & $^{+0.0018}_{-0.0012}$ \\
$R_{\rm P}^{2}$/R$_{*}^{2}$ & 0.01199 & $^{+0.00052}_{-0.00039}$ \\
$b$ $\equiv$ $a \cos i/R_{\rm *}$  & 0.798 & $^{+0.026}_{-0.019}$ \\
\\
$e$ & \multicolumn{2}{c}{0 (adopted)}\\
$K_{\rm 1}$ (km s$^{-1}$) & 0.1167 & $^{+0.0024}_{-0.0019}$ \\
$\gamma$ (km s$^{-1}$)    & $-1.93619$ & $^{+0.00021}_{-0.00023}$ \\
$a$ (AU)    & 0.0421 & $^{+0.0010}_{-0.0018}$ \\
$i$ (degs)  & 85.22 & $^{+0.27}_{-0.43}$ \\
\\
$M_{\rm *}$ (\msun) & 1.022 & $^{+0.074}_{-0.129}$ \\
$R_{\rm *}$ (\rsun) & 0.946 & $^{+0.057}_{-0.052}$ \\
$\log g_*$ (cgs)    & 4.495 & $^{+0.030}_{-0.054}$ \\
$\rho_*$ ($\rho_\odot$) & 1.21 & $^{+0.13}_{-0.18}$ \\
\\
$M_{\rm p}$ (\mjup) & 0.855 & $^{+0.043}_{-0.076}$ \\
$R_{\rm p}$ (\rjup) & 1.008 & $^{+0.083}_{-0.060}$ \\
$\rho_{\rm p}$ ($\rho_{\rm{Jup}}$) & 0.83 & $^{+0.13}_{-0.17}$ \\
$\log g_{\rm p}$ (cgs) & 3.284 & $^{+0.041}_{-0.064}$ \\
$T_{eq} (A=0, F=1)$ (K) & 1280 & $^{+35}_{-21}$ \\
Safronov number ($\Theta$) & 0.070 & $\pm0.010$ \\
\enddata
\end{deluxetable}

\subsection{Planet parameters}

The CORALIE spectroscopic RV data were combined with the WASP-South and EULERCAM
photometric data in a simultaneous fit to determine the planetary parameters.
The method of Markov Chain Monte Carlo (MCMC) as detailed in previous
investigations \citep{pollacco2008wasp3,cameron2007} was used. We use the
\citet{claret2000} limb darkening coefficients for the appropriate stellar
temperature and photometric passband and a adaptive stepsize mechanism is used
during the 5000 step burn-in phase until the chain converges. At the end of this
phase, the adaptive stepsize mechanism is switched off for the final 20000 steps
in the chain. The autocorrelation length of the chain was $9\pm1$ for all the
parameters, indicating that no unwanted correlations are present and the chain
is ``well-mixed''. 

Initial fits showed that the eccentricity was poorly constrained but consistent
with zero and so was fixed at this value in subsequent fits. The prior on the
stellar mass was set to 1.0\,\msun, as indicated by the evolutionary tracks
discussed in the previous section, but no constraint or prior on the stellar
radius or density was used in the fit. The transit parameters such as the
period, depth, duration were initially set at the values from the transit search
of the WASP-South data and subsequently refined in the MCMC code using all the
available data. 

The best fitting system parameters are listed in Table~\ref{tab:wasp16-sysparams}
and show that WASP-16b is a reasonably close Jupiter analog albeit somewhat less
massive and in a $P\sim3$\,day orbit. The host star has a fitted mass and radius
which are slightly smaller than the Sun, leading to a slightly higher density
than the solar case but all the parameters are identical to the Sun within the
error bars. The lack of lithium detection, low \vsini\  and similar large inferred age
also point towards WASP-16 being a solar analog hosting a hot Jupiter planet.

\section{Times of transit}

Although WASP-16 was observed with WASP-South for one full and two partial
seasons, there are very few complete transits within the timeseries suitable for
determining times of transits. This illustrates the need for long timeseries on
potential transit fields as shown by \citet{smith2006}. In total we find four
complete and well measured transits from the SuperWASP data and these are shown
in Table~\ref{tab:wasp16times} as \textit{`Fitted $T_o$'} along with one time of transit determined from
the EULERCAM data. The predicted times of transit from the MCMC ephemeris (given in
Table~\ref{tab:wasp16-sysparams}) are also shown in Table~\ref{tab:wasp16times}.
There is currently an insufficient number of measured transits with adequate
precision to suggest anything other than a constant period.

\begin{deluxetable}{cccc}
\tablecaption{Times of transit for WASP-16b.\label{tab:wasp16times}}
\tablehead{
\colhead{Fitted $T_0$} & \colhead{Error} & \colhead{Predicted $T_0$} & \colhead{O-C} \\
\colhead{(HJD-2450000)} & \colhead{(days)} & \colhead{(HJD-2450000)} & \colhead{(days)}}
\tablewidth{0pt}
\tablecolumns{4}
\startdata
4216.43288 & 0.00243 & 4216.43417 & -0.00129 \\
4238.26861 & 0.00458 & 4238.26436 & 0.00425 \\
4266.31992 & 0.06319 & 4266.33175 & -0.01183 \\
4291.27823 & 0.03342 & 4291.28054 & -0.00231 \\
4590.66606 & 0.00028 & 4590.66602 & 0.00004 \\
\enddata
\end{deluxetable}

\section{Conclusions}
We report the discovery of a new transiting planet with the WASP-South station
of the SuperWASP survey. The planet, designated WASP-16b, orbits a star which is
a close solar analog, having temperature, mass, radius, metallicity and gravity the same as
the Sun, within the error bounds. The age of the host star, with an admittedly
large error bar, is also close to the solar age.

The orbiting planet is a reasonable Jupiter analog although somewhat less
massive than Jupiter ($M_p\sim0.85$\,\mjup), but with a near identical radius
($R_p\sim1.01$\,\rjup) leading to a density some 80\% of Jupiter. This planet
falls in the lower left corner of the group of ``normal'' Jupiter-sized planets
in the Mass/Radius diagram, with the majority of objects in this region being
somewhat larger than Jupiter in either mass or radius. Additionally if we compute
the Safronov number $\Theta\equiv\frac{1}{2}(V_{esc}/V_{orb})^2 = 0.070\pm0.010$ 
for this planet, this along with it's equilibrium temperature $T_{eq}=1280$\,K
places it in the center of the Class I planets as defined by \cite{hansen2007}. 
The ``normality'' of this planet makes it similar to WASP-2b, TrES-1b and other
``normal'' extrasolar planets ans stands in contrast to the inflated radii and
low densities of planets like TrES-4b \citep{sozzetti2009}, HD~209458b
\citep{brown2001} and WASP-1b \citep{cameron2007}.

\acknowledgments
The WASP Consortium comprises the Universities of Keele, Leicester, St. Andrews,
the Queen's University Belfast, the Open University and the Isaac Newton Group.
WASP-South is hosted by the South African Astronomical Observatory and we are
grateful for their support and assistance. Funding for WASP comes from the
consortium universities and from the UK's Science and Technology Facilities
Council.

\bibliographystyle{apj}                       
\bibliography{apj-jour,wasp16}

\begin{thebibliography}{22}
\expandafter\ifx\csname natexlab\endcsname\relax\def\natexlab#1{#1}\fi

\bibitem[{Alonso {et~al.}(2004)Alonso, Brown, Torres, Latham, Sozzetti,
  Mandushev, Belmonte, Charbonneau, Deeg, Dunham, O'Donovan, \&
  Stefanik}]{alonso2004}
Alonso, R., Brown, T.~M., Torres, G., Latham, D.~W., Sozzetti, A., Mandushev,
  G., Belmonte, J.~A., Charbonneau, D., Deeg, H.~J., Dunham, E.~W., O'Donovan,
  F.~T., \& Stefanik, R.~P. 2004, \apj, 613, L153

\bibitem[{{Bakos} {et~al.}(2002){Bakos}, {L{\'a}z{\'a}r}, {Papp}, {S{\'a}ri},
  \& {Green}}]{bakos2002hat}
{Bakos}, G.~{\'A}., {L{\'a}z{\'a}r}, J., {Papp}, I., {S{\'a}ri}, P., \&
  {Green}, E.~M. 2002, \pasp, 114, 974

\bibitem[{{Baraffe} {et~al.}(1998){Baraffe}, {Chabrier}, {Allard}, \&
  {Hauschildt}}]{baraffe1998}
{Baraffe}, I., {Chabrier}, G., {Allard}, F., \& {Hauschildt}, P.~H. 1998, \aap,
  337, 403

\bibitem[{{Baranne} {et~al.}(1996){Baranne}, {Queloz}, {Mayor}, {Adrianzyk},
  {Knispel}, {Kohler}, {Lacroix}, {Meunier}, {Rimbaud}, \& {Vin}}]{baranne1996}
{Baranne}, A., {Queloz}, D., {Mayor}, M., {Adrianzyk}, G., {Knispel}, G.,
  {Kohler}, D., {Lacroix}, D., {Meunier}, J.-P., {Rimbaud}, G., \& {Vin}, A.
  1996, \aaps, 119, 373

\bibitem[{{Blackwell} \& {Shallis}(1977)}]{blackwell1977}
{Blackwell}, D.~E. \& {Shallis}, M.~J. 1977, \mnras, 180, 177

\bibitem[{{Brown} {et~al.}(2001){Brown}, {Charbonneau}, {Gilliland}, {Noyes},
  \& {Burrows}}]{brown2001}
{Brown}, T.~M., {Charbonneau}, D., {Gilliland}, R.~L., {Noyes}, R.~W., \&
  {Burrows}, A. 2001, \apj, 552, 699

\bibitem[{{Claret}(2000)}]{claret2000}
{Claret}, A. 2000, \aap, 363, 1081

\bibitem[{Collier~Cameron {et~al.}(2006)Collier~Cameron, Bouchy, {H\'ebrard}, Maxted, Pollacco,
  Pont, Skillen, Smalley, Street, West, Wilson, Aigrain, Christian, Clarkson,
  Enoch, Evans, Fitzsimmons, Fleenor, Gillon, Haswell, Hebb, Hellier, Hodgkin,
  Horne, Irwin, Kane, Keenan, Loeillet, Lister, Mayor, Moutou, Norton, Osborne,
  Parley, Queloz, Ryans, Triaud, Udry, \& Wheatley}]{cameron2006}
Cameron, A.~C et~al., 2006, \mnras, 373, 799
  
\bibitem[{Collier~Cameron {et~al.}(2007)Collier~Cameron, Bouchy, {H\'ebrard}, Maxted, Pollacco,
  Pont, Skillen, Smalley, Street, West, Wilson, Aigrain, Christian, Clarkson,
  Enoch, Evans, Fitzsimmons, Fleenor, Gillon, Haswell, Hebb, Hellier, Hodgkin,
  Horne, Irwin, Kane, Keenan, Loeillet, Lister, Mayor, Moutou, Norton, Osborne,
  Parley, Queloz, Ryans, Triaud, Udry, \& Wheatley}]{cameron2007}
Cameron, A.~C. et~al., 2007, \mnras, 375, 951


\bibitem[{{Crouzet} {et~al.}(2009){Crouzet}, {Agabi}, {Blazit}, {Bonhomme},
  {Fante{\"i}-Caujolle}, {Fressin}, {Guillot}, {Schmider}, {Valbousquet},
  {Bondoux}, {Challita}, {Abe}, {Daban}, \& {Gouvret}}]{crouzet2008}
{Crouzet}, N., {Agabi}, K., {Blazit}, A., {Bonhomme}, S.,
  {Fante{\"i}-Caujolle}, Y., {Fressin}, F., {Guillot}, T., {Schmider}, F.-X.,
  {Valbousquet}, F., {Bondoux}, E., {Challita}, Z., {Abe}, L., {Daban}, J.-B.,
  \& {Gouvret}, C. 2009, in Transiting Planets - Proceedings of IAU Symposium
  No. 253, ed. F.~Pont, D.~Queloz, \& D.~Sasselov (Cambridge University Press),
  336--339

\bibitem[{Girardi {et~al.}(2000)Girardi, Bressan, Bertelli, \&
  Chiosi}]{girardi2000}
Girardi, L., Bressan, A., Bertelli, G., \& Chiosi, C. 2000, \aaps, 141, 371

\bibitem[{Hansen \& Barman(2007)}]{hansen2007}
Hansen, B.~M.~S. \& Barman, T. 2007, \apj, 671, 861

\bibitem[{McCullough {et~al.}(2005)McCullough, Stys, Valenti, Fleming, Janes,
  \& Heasley}]{mccullough2005}
McCullough, P.~R., Stys, J.~E., Valenti, J.~A., Fleming, S.~W., Janes, K.~A.,
  \& Heasley, J.~N. 2005, \pasp, 117, 783

\bibitem[{{Pollacco} {et~al.}(2008){Pollacco}, {Skillen}, {Collier Cameron},
  {Loeillet}, {Stempels}, {Bouchy}, {Gibson}, {Hebb}, {H{\'e}brard}, {Joshi},
  {McDonald}, {Smalley}, {Smith}, {Street}, {Udry}, {West}, {Wilson},
  {Wheatley}, {Aigrain}, {Alsubai}, {Benn}, {Bruce}, {Christian}, {Clarkson},
  {Enoch}, {Evans}, {Fitzsimmons}, {Haswell}, {Hellier}, {Hickey}, {Hodgkin},
  {Horne}, {Hrudkov{\'a}}, {Irwin}, {Kane}, {Keenan}, {Lister}, {Maxted},
  {Mayor}, {Moutou}, {Norton}, {Osborne}, {Parley}, {Pont}, {Queloz}, {Ryans},
  \& {Simpson}}]{pollacco2008wasp3}
{Pollacco}, D. et~al. 2008, \mnras, 385, 1576

\bibitem[{Pollacco {et~al.}(2006)Pollacco, Skillen, Collier~Cameron, Christian,
  Irwin, Lister, Street, West, Anderson, Clarkson, Deeg, Enoch, Evans,
  Fitzsimmons, Haswell, \& {13 others}}]{pollacco2006}
Pollacco, D.~L., et~al. 2006, \pasp, 118, 1407

\bibitem[{{Queloz} {et~al.}(2001){Queloz}, {Henry}, {Sivan}, {Baliunas},
  {Beuzit}, {Donahue}, {Mayor}, {Naef}, {Perrier}, \& {Udry}}]{queloz2001}
{Queloz}, D., {Henry}, G.~W., {Sivan}, J.~P., {Baliunas}, S.~L., {Beuzit},
  J.~L., {Donahue}, R.~A., {Mayor}, M., {Naef}, D., {Perrier}, C., \& {Udry},
  S. 2001, \aap, 379, 279

\bibitem[{Sestito \& Randich(2005)}]{sestito2005}
Sestito, P. \& Randich, S. 2005, \aap, 442, 615

\bibitem[{Smith {et~al.}(2006)Smith, Cameron, Christian, Clarkson, Evans,
  Haswell, Hellier, Horne, Irwin, Kane, Lister, Norton, Pollacco, Skillen,
  Street, Triaud, West, \& Wheatley}]{smith2006}
Smith, A.~M.~S. et~al.  2006, \mnras, 373, 1151

\bibitem[{{Sozzetti} {et~al.}(2009){Sozzetti}, {Torres}, {Charbonneau}, {Winn},
  {Korzennik}, {Holman}, {Latham}, {Laird}, {Fernandez}, {O'Donovan},
  {Mandushev}, {Dunham}, {Everett}, {Esquerdo}, {Rabus}, {Belmonte}, {Deeg},
  {Brown}, {Hidas}, \& {Baliber}}]{sozzetti2009}
{Sozzetti}, A. et~al. 2009, \apj, 691, 1145

\bibitem[{{Strassmeier} {et~al.}(2007){Strassmeier}, {Andersen}, {Granzer},
  {Korhonen}, {Herber}, {Cutispoto}, {Rafanelli}, \&
  {Horne}}]{strassmeier2007icet}
{Strassmeier}, K.~G., {Andersen}, M.~I., {Granzer}, T., {Korhonen}, H.,
  {Herber}, A., {Cutispoto}, G., {Rafanelli}, P., \& {Horne}, K. 2007, in ASP
  Conference Series, Vol. 366, Transiting Extrapolar Planets Workshop, ed.
  C.~{Afonso}, D.~{Weldrake}, \& T.~{Henning}, 332--334

\bibitem[{{Torres} {et~al.}(2005){Torres}, {Konacki}, {Sasselov}, \&
  {Jha}}]{torres2005}
{Torres}, G., {Konacki}, M., {Sasselov}, D.~D., \& {Jha}, S. 2005, \apj, 619,
  558

\bibitem[{{West} {et~al.}(2009){West}, {Anderson}, {Gillon}, {Hebb}, {Hellier},
  {Maxted}, {Queloz}, {Smalley}, {Triaud}, {Wilson}, {Bentley}, {Collier
  Cameron}, {Enoch}, {Horne}, {Irwin}, {Lister}, {Mayor}, {Parley}, {Pepe},
  {Pollacco}, {Segransan}, {Udry}, \& {Wheatley}}]{west2009wasp15}
{West}, R.~G. et~al. 2009, \aj, 137, 4834

\end{thebibliography}

{\it Facilities:} \facility{Euler1.2m}, \facility{ESO:3.6m}

\clearpage


\end{document}